# Towards Earth AntineutRino TomograpHy (EARTH)

R.J. de Meijer[1,2]), F.D. Smit[3]), F.D. Brooks[4]), R.W. Fearick[4]), H.J. Wörtche[5]) and F. Mantovani[6])

[1]) Stichting EARTH, 9321 XS2, Peize, the Netherlands

[2]) Dept. of Physics, University of the Western Cape, Belleville, South Africa.

[3]) iThemba LABS, PO Box 722, Somerset West 7129, South Africa

[4]) Dept. of Physics, University of Cape Town, Private Bag, Rondebosch 7701, South Africa

[5]) Rijksuniversiteit Groningen, 9744AA, Groningen, the Netherlands

[6]) Centro di GeoTecnologie, Siena University, Via Vetri Vecchi 34, I-52027, San Giovanni Valdarno, Arezzo, Italy

Corresponding author: R.J. de Meijer, de Weehorst, 9321 XS Peize, the Netherlands, phone: +31-50-5016654; email: demeijer@geoneutrino.nl

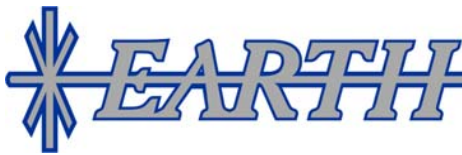

**Abstract.**
The programme Earth AntineutRino TomograpHy (EARTH) proposes to build ten underground facilities each hosting a telescope. Each telescope consists of many detector modules, to map the radiogenic heat sources deep in the interior of the Earth by utilising direction sensitive geoneutrino detection. Recent hypotheses target the core-mantle boundary (CMB) as a major source of natural radionuclides and therefore of radiogenic heat. A typical scale of the processes that take place at the CMB is about 200km. To observe these processes from the surface requires an angular resolution of about 3˚. EARTH aims at creating a high-resolution 3D-map of the radiogenic heat sources in the Earth's interior. It will thereby contribute to a better understanding of a number of geophysical phenomena observed at the Earth's surface. This condition requires a completely different approach from the monolithic detector systems as e.g. KamLAND.

This paper presents, for such telescopes, the boundary conditions set by physics, the estimated count rates, and the first initial results from Monte Carlo simulations and laboratory experiments. The Monte Carlo simulations indicate that the large volume telescope should consist of detector modules each comprising a very large number of detector units, with a cross section of roughly a few square centimetres. The signature of an antineutrino event will be a double pulse event. One pulse arises from the slowing down of the emitted positron, the other from the neutron capture. In laboratory experiments small sized, $^{10}$B-loaded liquid scintillation detectors were investigated as candidates for direction sensitive, low-energy antineutrino detection.

# 1. Introduction.

*How does the Earth's Interior work?*
In its special issue Science in July 2005 listed this question as one of the 25 most prominent questions for the next 25 years (Kerr, 2005). In October 2005 the Scientific American produced a special edition on "Our Ever Changing Earth". This indicates a revitalisation of widespread interest in the interior of our planet. At first glance this sounds surprising. It seems that in contrast to the successful exploration of our Solar system and parts of the Universe, we have a very limited knowledge of the interior of our planet. The deepest that has been drilled into the Earth was ~13km deep, a mere 0.1% of the Earth diameter and corresponding to the cruising altitude of jetliners. With present techniques, to "descending" deeper is prevented by a rapid increase in temperature and pressure.

Since the beginning of the 20$^{th}$ century information on the deeper parts has been derived from the speed, reflection and refraction of seismic waves, the moment of inertia and the precession motion of the planet, and the physical, chemical and mineralogical information obtained from meteorites and xenolithes. Our present knowledge is often schematically in spherical symmetric models having a crust floating on a viscous mantle, subdivided into a number of concentric shells and encompassing a partially liquid core (Oldham, 1906; Gutenberg, 1914). Only in the crust and the upper mantle usually some structure is indicated.

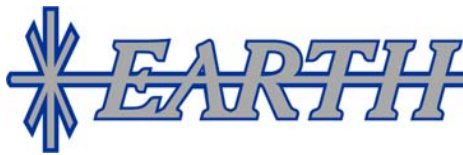

In the last decades of the 20th century through developments in seismic tomography it has been revealed that parts of the crust are being subducted and have reached the deeper parts of the mantle. The previous view that the convective flow is stratified at a depth of 660 km and an unmixed or pristine lower mantle is preserved is no longer tenable (Van der Hilst and Karson, 1999; Zhao, 2004).

Boyet and Carlson(2005) present a new view on the Earth's interior, which is based on the differences in the isotopic abundance of $^{142}$Nd found in meteorites and mantle-derived terrestrial samples. One of the new features is that the layer at the core-mantle boundary (CMB) is a reservoir enriched in radiogenic heat producing sources, resulting from a distillation of the earlier magma ocean and subducted crust. This layer is likely to be the origin of the deep volcanic plumes that manifest themselves at the Earth's surface as ocean islands (e.g. Hawaii, Iceland, Galapagos and Curaçao).

Wilson(2005) quoting Tolstikhin and Hoffman(2005) speculates that this 'hidden' reservoir is composed out of the ancient primordial crust formed from the solidifying magma ocean. Regardless of its precise dimensions and location, the hidden reservoir is thought to contain over 40% of Earth's K, Th and U, the main heat producing elements. If it resides on the core-mantle boundary, the layer would form a blanket of heat, consistent with the temperature jump of 1000 to 2000 K within a few hundred kilometres as proposed by Lay *et al.*(1998).

Presently, little detail is known regarding the fate of subducting slabs. It is clear that large earthquakes occur at the slab/continent interface, and within the slab down to about 670 km depth. But just how far the slab penetrates into the mantle and how the rheology of lithospheric materials behaves once they reach the lower mantle are currently matters of active debate. These questions are fundamental issues in Earth Sciences since they relate to the nature of mantle convection as well as how the Earth evolved and cools off: Or, in other words, How the Earth's Interior "works"?

Processes in the deeper Earth manifest themselves at the surface. The convection in the liquid core produces the geomagnetic field, while the convection in the mantle leads to drift of ocean plates and continents as well as volcanic plumes forming ocean islands. These processes are driven by heat flow. The location, type and size of the heat sources are still a topic of debate. We know the heat flow at the Earth's surface from measurements at about 25000 locations. These measurements have lead to a rather detailed heat-flow map (Pollack *et al.*, 1993). The map shows a large variation (factor 20) in heat flow at the surface with maxima at the mid-oceanic ridges on the southern hemisphere. Integrating the mapped yields, gives a total heat flow of about 45 TW, which is equivalent to the heat production of 15000 power plants of 1000MW$_e$ each.

*Geoneutrinos.*
According to Buffett, 6-12TW is produced in the crust (2003) and is of radiogenic origin (decay of natural radionuclides). Radiogenic processes are considered to be predominantly responsible for also the remaining part of the heat flow. In addition to heat the radiogenic processes in the Earth produce antineutrinos and neutrinos: they have been named geoneutrinos (Araki *et al.*, 2005).

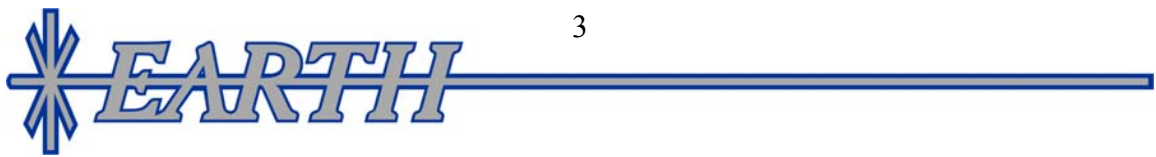

The heat produced in nuclear decay is directly related (Fiorentini *et al.*, 2003) to the flux of antineutrinos, as is illustrated in Table 1. Detection of low-energy antineutrinos produced in the U and Th decay processes has been demonstrated by liquid scintillator detectors in Kamioka, Japan and in Chooz and Bugey-3 in France. These set-ups primarily address the fundamental aspects of antineutrinos such as their flavour changes and the related mixing angles. The principle of these detectors is based on the capture of an antineutrino, $\bar{\nu}$, by a proton of the scintillator material, producing a positron and a neutron. In a simplified picture the positron carries the energy information and the neutron is emitted preferentially in the same direction as the incoming antineutrino (Beacom and Vogel, 1996). The neutron travels only a few centimetres before it is captured. In these detectors the neutron capture is detected by the emitted capture γ-ray. The delayed coincident detection between positron emission and neutron capture characterises an antineutrino detection signature.

The KamLAND collaboration published the first official results on the detection of geoneutrinos (Araki *et al.*, 2005) in July 2005. Their results were obtained with a 1 kiloton detector, filled with liquid scintillator and housed in an underground mine near Kamioka, Japan. The detector was originally designed for the detection of fundamental properties of antineutrinos and for this reason is located in the vicinity of nuclear power plants. The geoneutrinos therefore are superimposed on a bell-shaped continuum ranging up to about 8 MeV in the antineutrino spectrum caused by antineutrinos resulting from fission processes in the power reactor. The KamLAND detector is a monolithic liquid scintillator detector with a mass of about 1 kilotonne. The data presented by Araki *et al.* (2005) correspond to a measuring period of about 750 days. They have been analysed after making extensive corrections for antineutrinos from the power plants and spurious events due to cosmic-ray induced reactions such as $^{13}C(\alpha,n)^{16}O$, which mimicke geoneutrino events.

## 2. Proposed geoneutrino telescope.
The need for high-resolution antineutrino tomography to map the radiogenic heat sources in the Earth's interior has set the goals for the Earth AntineutRino TomograpHy (EARTH) programme, initially presented in 2004 (de Meijer *et al.*, 2004a,b). To resolve structures of about 150km diameter in the CMB requires an angular resolution of about 3˚. This goal is to be realised by a set of ten telescopes distributed worldwide, each with a resolution of about 10˚ achieved by using direction sensitive detector modules. This goal sets a number of boundary conditions that more or less dictates our starting point and the initial direction of our technological research. We are fully aware that our task is ambitious and not straightforward and cannot be achieved with state of the art technology. Hence it requires a step-wise approach with manageable tasks, clear deliverables and go/no-go decisions. It seems feasible, but requires considerable technological development, with therefore, in all likelihood quite a number of spin-offs.

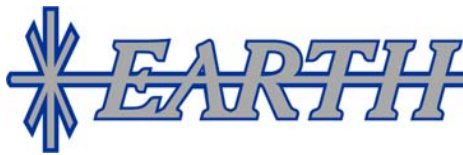

The results of KamLAND confirm the feasibility of geoneutrino detection by large volume detectors, but with the present monolithic detectors, no location of the geoneutrino sources can be made. As indicated above, the antineutrino capture contains information on the direction of the incoming antineutrino and the challenge becomes how to utilise this information. As we will demonstrate in this proposal for the localisation of the radiogenic heat sources, directional sensitive detectors are to be developed. These detectors will be placed in a modular detector set-up to form a telescope with a detector mass of four times that of KamLAND.

To check the feasibility and the degree of directional discrimination we place our first telescope, TeleLENS (Telescope for Low-Energy Neutrino based Sciences), on the island of Curaçao, Netherlands Antilles, situated at about 12˚N; 69W˚. Using the crustal reference model as used by Mantovani *et al.*, (2004) and assuming 20TW homogeneously distributed in the mantle as well as a localised hypothetic source of 5TW in the CMB at 30˚S; 69W˚, we estimate an expected signal of 24 TNU from the continental and oceanic crusts, 17 TNU from the mantle and 6 TNU from the hypothetical localised source. TNU stands for Terrestrial Neutrino Unit and corresponds to one antineutrino event per year and per $10^{32}$ proton. Since the conversion to the number of detected geoneutrinos depends on the actual volume of the detectors, the detector material and the detection efficiency, it is hard to produce reliable numbers for TeleLENS at this stage of its development.

The choice for Curaçao comes from its large distance to operating nuclear power stations in Florida. The ratio between the fluxes of antineutrinos from the power reactors and geoneutrinos at Curaçao is 0.1. This value is similar to the ratio for Hawaii and two orders of magnitude smaller than for Kamioka, Japan. According to geological information obtained from surface studies (Beets, 1972; Klaver, 1987) the western part of the island contains a large and deep body of limestone on top of basalt. Plans for an analysis of seismic data and pilot drilling are presently being discussed, partly in the framework of the International Scientific Drilling Programme (ICDP). A first exploratory drill to 70m depth revealed a temperature drop with depth similar to that in the ocean waters.

Direction sensitive antineutrino detectors have not yet successfully been demonstrated and therefore the design, construction and test of these detectors will be one of the objectives of the first phase of the EARTH programme. At this time, our prime emphasis is on the development of the detector units and modules as well as their associated electronics and read-out systems. This development should lead to a Proof of Principle test of our direction sensitive detectors, planned to be carried out at the nuclear power plant of Koeberg, ~25km north of Cape Town. The outcome of this test is the first go/no-go decision point. We therefore refrain from further speculations on the subsequent trajectories or the details of TeleLENS.

## 2.1 Technological approach
Antineutrino detection is traditionally based on the inverse β-decay:

$$\bar{\nu}_e + p \rightarrow n + e^+$$

in which an antineutrino, $\bar{\nu}_e$, is captured by a proton producing a positron and a neutron. The reaction has a Q-value of -1.8MeV, hence in a scintillator, geoneutrinos produced by $^{40}$K are not detected (see table 1). As mentioned above, in a simplified picture the positron carries the energy information, and the neutron is emitted preferentially in the same direction as the incoming antineutrino. The neutron is detected via charged particles or photons emitted directly after it is captured by a nucleus in the scintillator material. The neutron travels a few centimetres in a few microseconds before it is captured. The delayed coincident detection between the positron and the neutron defines an antineutrino detection signature.

Table 1. Maximum (anti)neutrino energy and heat production in natural decay processes.

| Decay | $E_{max}$(MeV) | Heat(W/kg) |
|---|---|---|
| $^{238}U \rightarrow ^{206}Pb + 8\,^4He + 6e + 6\bar{\nu}$ | 3.26 | $0.95 \cdot 10^{-4}$ |
| $^{232}Th \rightarrow ^{208}Pb + 6\,^4He + 4e + 4\bar{\nu}$ | 2.25 | $0.27 \cdot 10^{-4}$ |
| $^{40}K \rightarrow ^{40}Ca + e + \bar{\nu}$ (88.8%) | 1.31 | $0.36 \cdot 10^{-8}$ |
| $^{40}K + e \rightarrow ^{40}Ar + \nu$ (11.2%) | 1.51 | |

Traditionally the neutron capture takes place on a H or a Gd nucleus within the scintillator. Prompt γ-rays resulting from neutron capture are detected. The γ-ray emission is isotropic and the mean free path of the γ-rays is considerably larger than the few centimetres the neutron travels. Hence, the direction information carried by the neutron is lost. In our proposal $^{10}$B is used as a neutron catcher. Capture of a thermal or epithermal neutron on $^{10}$B leads to disintegration into two charged particles (α and $^7$Li nucleus) which are then brought to rest in the scintillator within a few microns from the point of capture. In addition, the energy-dependence of the neutron capture cross section ($1/v_n$) of $^{10}$B leads to an earlier capture of the neutron. This narrows the time window of the event and ensures that the neutron does not deviate too much from its original direction. Our Monte Carlo simulations (see below) show a reduction in the number of collisions (between positron emission and neutron capture) by a factor of two in scintillators containing 5 % (by weight) $^{10}$B compared with those containing no boron.

Wang *et al.* (1999) have discussed the feasibility of using boron-loaded liquid scintillators for the detection of antineutrinos. Based on their work together with earlier studies of boron-loaded scintillators (BLS) we consider that this detection medium is not suitable for use in large scale monolithic detectors such as KamLAND but could nevertheless be useful in a large detector system consisting of a large number of relatively small (< 10 litre) detector modules. Two important factors that prohibit the use of BLS in a large monolithic antineutrino detector are the following: (a) only a liquid BLS could be considered for use in such a large detector and the liquid BLS presently available are all highly hygroscopic and would thus be extremely difficult to handle and to contain in large volumes; and (b) even though the kinetic energy released to the charged products of the neutron capture reaction in BLS is >2.3 MeV the light output resulting from this energy is very small, equivalent to that produced by an electron of energy about 60 keV, due to the well-known ionization density quenching characteristics of organic scintillators. In a very large scintillator the light attenuation due to the long travel distance through the scintillator to the photomultiplier tubes can be expected to reduce the weak neutron capture signal to a level at which it cannot be distinguished from photomultiplier noise and low energy background.

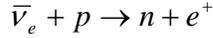

These problems can be avoided in a modular system if the design of a basic single detector unit is very simple (see for example Figure 2) and limited to a maximum volume of about 10 litres. In such a system there is first the possibility of using a plastic BLS which is rugged and chemically stable, unlike the liquid BLS. However, even if a liquid BLS is used handling and containment of the liquid should not be a problem when the volume is small. In addition, the good light collection properties that can be achieved using the proposed size of module should avoid problems in the detection of the small amplitude of the neutron capture signal in BLS.

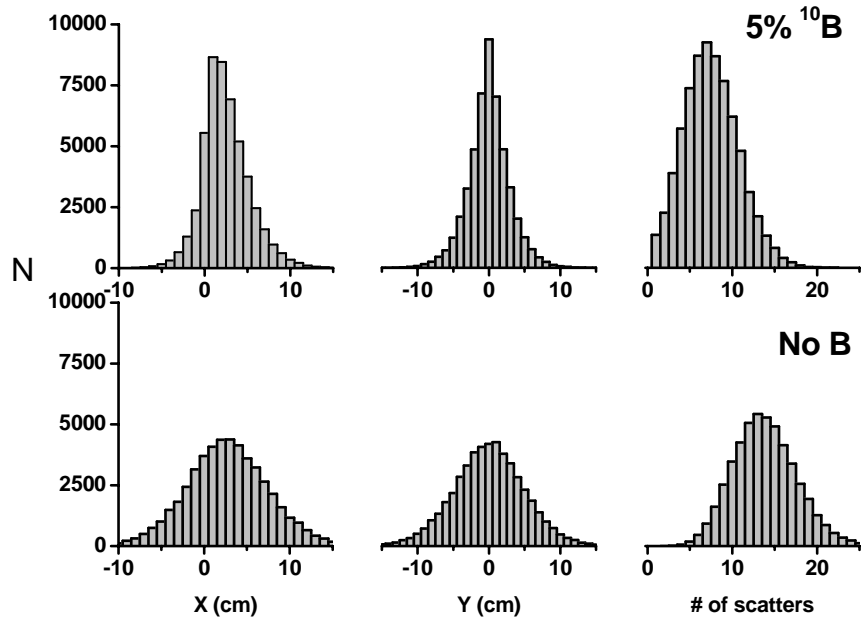

Fig.1. The effect of $^{10}$B loading on the neutron capture location. At the top a detector loaded with 5% $^{10}$B; at the bottom a detector without B. The results represent 50000 antineutrino capture simulations of reactor antineutrinos coming in along the negative x-axis and captured at (x,y)=(0,0) in a large volume detector.

*2.2 Simulations*

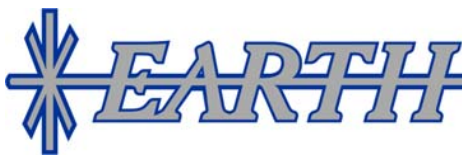

The influence of $^{10}$B on the direction information carried by the neutron has been investigated by simulating the capture reaction of an antineutrino by a proton according to the kinematics as described by Beacom and Vogel (1996). Figure 1 clearly shows the effect of $^{10}$B on the longitudinal and transverse distribution of the position where the neutron is captured. It clearly shows that the transverse distribution is much narrower and the longitudinal distribution is also considerably more focussed. These effects are clearly related to the fact that the neutron, on average, has half the number of collisions before it is captured.

These results have been used to simulate, in a single detector unit, the sensitivity of the detection probability for neutron detection to the angle of incidence of the antineutrino. For simplicity the unit is assumed to be very long relative to its cross section (two units are schematically presented in figure 2).

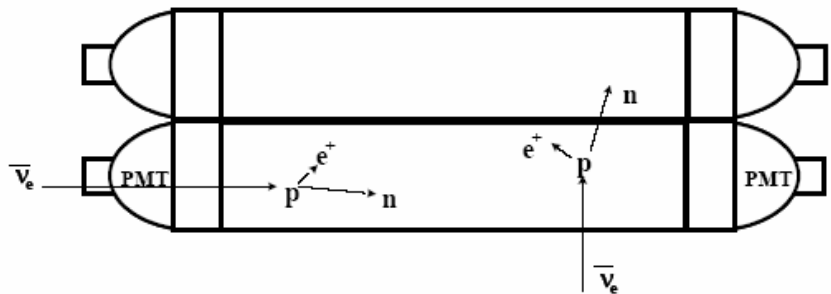

Fig. 2. Schematic view of two detector units in an antineutrino modules.

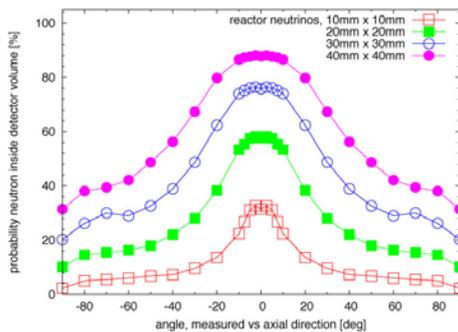

Fig. 3. Neutron-detection probability as function of incident angle and detector diameter.

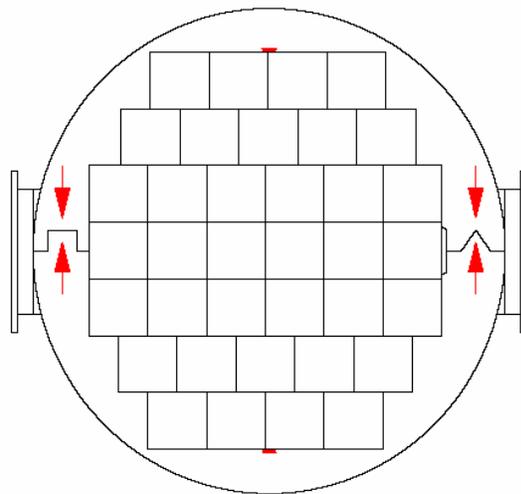

Fig. 4. Schematic representation of a multi-unit antineutrino module.

Figure 3 shows our simulated neutron detection probability as function of incident antineutrino angle relative to the detector axis for various detector cross sections. It clearly shows that direction sensitivity can only be obtained for small cross section, modular detector systems. It also shows that the efficiency of a single-unit detector is directly proportional to its cross section, indicating that without appropriate measures events will be discarded. With the low cross section for antineutrino capture this will be unacceptable. To solve this problem detectors are stacked in modules as e.g. in figures 2 and 4. Neutrons arising from radially entering antineutrinos would be lost from a single detector module, but can be recorded in an adjacent detector.

The simulations confirm the fact that the neutron travels only a few centimetres, which dictates that for direction sensitivity, the cross section of the detectors should be restricted to a few centimetres or less. Consequently, to obtain a large volume implies that a very large number of units is imperative.

*2.3 Experimental tests*

To test the feasibility of our approach we have started to carry out some experiments at iThemba LABS, South Africa. In these experiments the detection mechanism is mimicked by using neutrons from a $^{252}$Cf spontaneous fission source, and 3.8 cm diameter, 2.5cm long, sealed glass cells filled with NE311A liquid scintillator containing 1 or 5% $^{10}$B by mass. The neutrons elastically scatter off protons and produce a recoil-proton scintillation that simulates the positron emitted in antineutrino capture. A double-pulse event results if the neutron is moderated after multiple scatterings and eventually captured by a $^{10}$B nucleus in the BLS. The test experiments were carried out using a 8-bit, digital sampling oscilloscope to digitize the photomultiplier-output pulse shapes and a desktop PC to read out, record and analyse the digital output information.

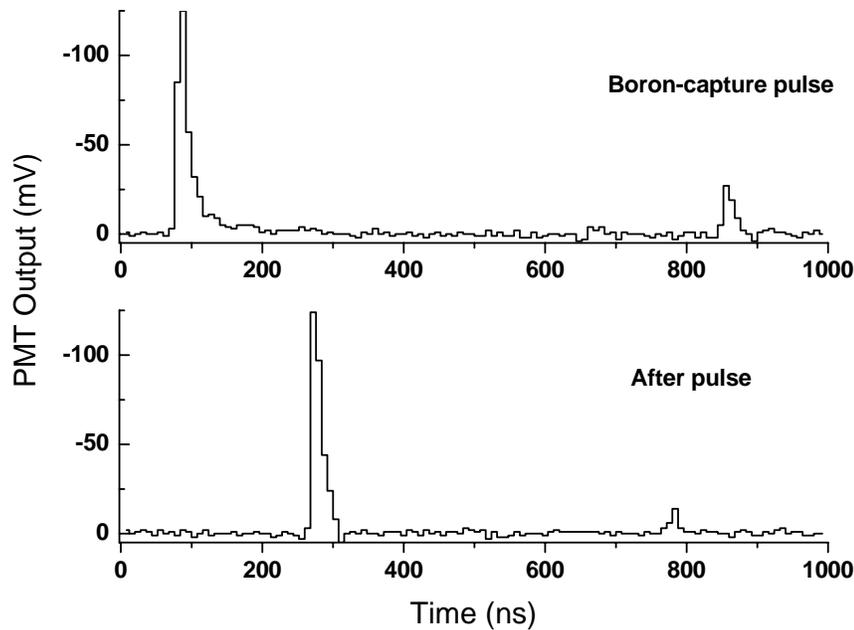

Figure 5. An example of two types of a double pulse event. Top: a recoil proton and a boron-capture pulse and bottom: a γ-ray pulse and an after pulse.

Figure 5 shows examples of two types of a double pulse event recorded in the experiments. Figure 5a shows a typical true double-pulse event. The initial pulse can be recognised as due to a recoil proton from the fact that it displays a distinct low-amplitude tail (slow scintillation component) that continues for 200-300 ns after the start of the pulse. The second pulse, due to neutron capture, stands out clearly above the background noise. Figure 5b shows an example of a spurious double-pulse event that can occur very easily in this type of detector and therefore needs to be well understood and carefully avoided. The event shown in this figure was obtained using a $^{60}$Co gamma source. Similar results can be obtained using any type of source, including both neutron and gamma. The initial pulse in figure 5b is attributed to a recoil electron associated with Compton scattering in the BLS. The "tail" of this pulse is small in comparison with that of the initial pulse in figure 5a and the two pulses can easily be distinguished as due to "electron" and "proton" respectively by means of a pulse-shape discrimination algorithm operating on the digital output data. The second pulse in figure 5b is attributed to "after-pulsing" associated with ion and/or optical feedback effects inside the photomultiplier tube of the detector. The after-pulse occurs at a characteristic time after the initial pulse (about 480 ns in the test experiments), depending on the operating conditions. It can be suppressed or controlled by careful selection of the high voltage applied to the photomultiplier and selection of the photomultiplier itself.

Figure 6 shows results from test measurements made under conditions in which after-pulsing was suppressed. The frequency of double-pulse events produced by neutrons from the $^{252}$Cf source was measured as a function of the time delay $T$ between the two pulses. The plot shows the number $N(T)$ of events for which this delay exceeds $T$ as a function of $T$. From simple considerations this distribution is expected to drop off exponentially with a decay time $T_0$ that depends on the concentration of $^{10}$B in the liquid scintillator, the detector geometry and perhaps other factors as well. Monte-Carlo simulations are in progress to determine $T_0$ for comparison with the experimental measurements.

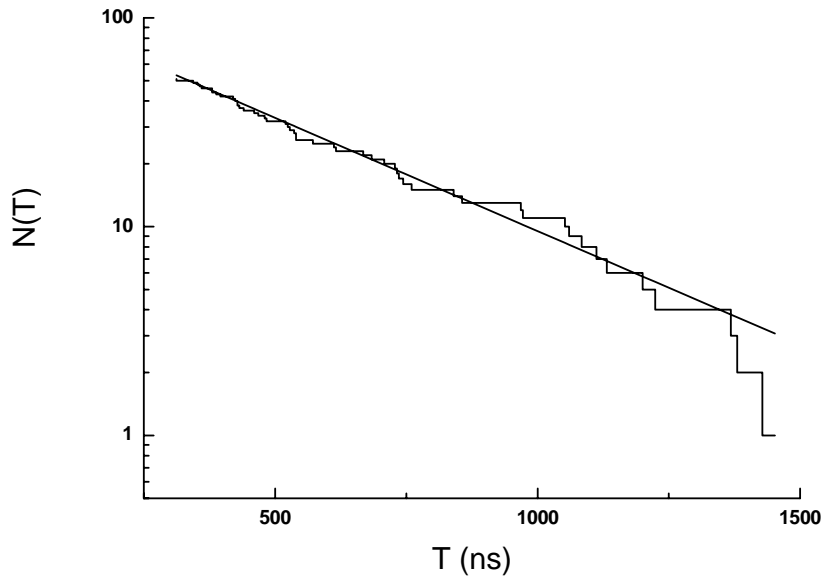

Figure 6. Plot of the number $N(T)$ of double pulse events for which the time between the two pulses is larger than $T$. The fitted line represents the function $116 \exp (T/T_0)$ with $T_0 = 400$ns.

## *2.4 Background reduction.*

One of the challenges for monolithic detectors is to reduce background; for modular detectors a similar challenge will be faced. In comparison to the monolithic detectors the modular detector is expected to have advantages in background reduction. In a monolithic detector every light producing event is detected by every PMT, if the signal is above its threshold. For a modular detector system the following factors contribute to background reduction:

1. The light pulse is only detected in a single cell, which comprises only a very small fraction of the total volume. (Estimated reduction factor: $10^{-6}$-$10^{-7}$)
2. A real event is restricted to one or a few adjacent cells.
3. The neutron capture by $^{10}$B produces an almost constant light pulse due to the large Q-value.
4. The close proximity of the PMTs (<1m) to the interaction leads to a higher light collection, which has two significant advantages
   - Capture on $^{10}$B produces a weaker neutron signature than capture on H or Gd. The close proximity may still allow the detection of the weaker signals. Moreover $^{10}$B loading leads to a faster capture of the neutron thereby better conserving the direction information carried by the neutron and reducing the interaction in space ($10^{-1}$-$10^{-2}$) and time ($10^{-6}$).

- Using the pulse characteristics becomes possible unlike in monolithic detectors.

Quantification of these background-reduction factors will be part of the Proof of Principle test.

*2.5 Proof of Principle*

The results obtained thus far give us confidence to proceed to the next stage on the route to the Proof of Principle test. In the next step we will first investigate the properties of boron-loaded plastic detectors as well as the use of natural boron-loaded liquid scintillators. These detectors will then be exposed to the high antineutrino flux at one of the Koeberg reactors (0.92 $GW_e$), located ~25km north of Cape Town. Based on the estimates of Bernstein *et al.*[21]) we expect about 2 to3 events per day per kilogram detector material. Initially we will mainly be interested in detection of double pulses and analysing the scintillator properties. After optimising the detectors and their electronics, we will construct a number of test detectors and investigate their individual direction sensitive detection efficiency.

## 3.Conclusions

A 3D image of the radiogenic heat sources in the Earth's interior with a spatial resolution of about 150km at the Core-Mantle-Boundary (CMB) will certainly revolutionise the understanding of how the Earth works and may lead to better knowledge on a number of phenomena observed at the Earth surface. In addition to seismic tomography, antineutrino tomography seems to be the only additional method to reach this goal. To obtain antineutrino tomography with a spatial resolution comparable or better than seismic tomography requires direction sensitive antineutrino detection. The existing large monolithic antineutrino detector set-ups will not be able to provide sufficient resolution.

The proposed detector system in this paper is a consequence of the goal to eventually map the radiogenic heat sources with high resolution by antineutrino tomography (e.g. located at the CMB with a size of about 200km). It starts by exploiting the direction information contained in the kinematics of the antineutrino capture by a proton. Based on simulations of the neutron tracking we conclude that the detectors should have a cross section of the order a few centimetre squared. In this paper we have demonstrated that from the physics point of view such detectors provide sufficient direction sensitivity and sufficient background reduction. In this paper we have not yet addressed the technical question of how a very large number of detectors may be equipped and read out. The presentation of Daniel Ferenc at this conference is an indication that with time these technical challenges can be resolved. We are fully aware that this is an indication of a solution to only one of the many technical developments that need to take place. On the one hand there is no guarantee of success but on the other hand we see no other obvious solution.

**References.**

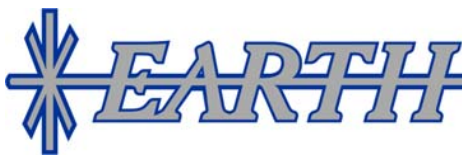